\documentclass[twocolumn,showpacs,showkeys,aps]{revtex4}
\usepackage{graphicx,amsmath,amsfonts,amssymb}
\newcommand{\be}{\begin{equation}}
\newcommand{\ee}{\end{equation}}
\newcommand{\wb}{\omega_b}
\newcommand{\bsup}[2]{\raisebox{#1 ex}{${\mbox{Sup }}\atop{#2}$}\ }
\def\Z{\mathbb{Z}}
\def\R{\mathbb{R}}

\date{February 3, 2004}
\begin{document}

\title{Bright and dark breathers in Fermi-Pasta-Ulam lattices}
\author{B. S\'anchez-Rey\dag, G. James\ddag, J. Cuevas\dag, and JFR Archilla\dag}
\address{\dag Nonlinear Physics Group, University of Seville, Spain \\ \ddag
Laboratoire Math\'ematiques pour l'Industrie et la Physique (UMR 5640),\\ INSA
de Toulouse, 135 avenue de Rangueil, 31077 Toulouse Cedex 4, France. }

\begin{abstract}
In this paper we study the existence and linear stability
of bright and dark breathers in one-dimensional FPU lattices. On the one hand,
we test the range of validity of a recent breathers existence proof [G. James,
{\em C. R. Acad. Sci. Paris}, 332, Ser. 1, pp. 581 (2001)] using numerical
computations. Approximate analytical expressions for small amplitude bright and
dark breathers are found to fit very well exact numerical solutions even far
from the top of the phonon band. On the other hand, we study numerically large
amplitude breathers non predicted in the above cited reference. In particular,
for a class of asymmetric FPU potentials we find an energy threshold for the
existence of exact discrete breathers, which is a relatively unexplored
phenomenon in one-dimensional lattices. Bright and dark breathers superposed on
a uniformly stressed static configuration are also investigated.
\end{abstract}
\pacs{63.20.Pw,63.20.Ry}
\keywords{intrinsic localized modes,fpu lattices}

\maketitle


\section{Introduction and model}
    Discrete breathers, also called intrinsic localized modes, are
classical exact spatially localized time-periodic solutions which can be
sustained by many nonlinear lattices (see \cite{Willis,SiPage} for a
review). In 1994 MacKay and
Aubry~\cite{Mackay} rigorously proved their existence in Hamiltonian lattices
with anharmonic on-site potentials and  weak  coupling. Breathers are obtained
by continuation from the uncoupled case in which trivial breathers exist (e.g.
breathers with only one oscillator excited, the others being at rest). With the
same technique, the existence of breathers was also proved for diatomic
Fermi-Pasta-Ulam (FPU) chains~\cite{Livi}. In this model, two different masses
alternate on the chain and are nonlinearly coupled to their nearest-neighbours
via an interaction potential $V$. This result is valid for a large mass ratio,
since light masses (next-nearest neighbours) are weakly coupled due to the
presence of a heavy mass in between.

Unfortunately this method is not applicable to homogeneous FPU lattices, which
do not possess an uncoupled limit in which trivial breathers exist. For some
years the only existence result concerned the particular potentials
$V(x)=x^{2m}$ with $m\ge 2$~\cite{Flach}. FPU lattices seemed elusive to a
rigorous mathematical treatment in spite of numerous
approximate~\cite{ST88,Pag90,Huang} and numerical~\cite{Bickham,Sandusky}
studies, which have indicated the existence of discrete breathers in these
systems.

    Nevertheless, recent papers have presented
rigorous existence proofs for discrete breathers in infinite FPU lattices. On
the one hand, Aubry et al~\cite{AKK01} have proved the existence of breathers
with frequencies above the  phonon spectrum, when $V$ is a strictly convex
polynomial of degree $4$. These results are obtained via a variational method
and apply in fact to higher dimensional generalizations of  FPU lattices.
Without additional assumptions on $V$, these results give only a partial
information on breathers amplitudes. Under the additional condition that $V$ is
even, Aubry et al prove the existence of breathers of arbitrarily small
amplitudes in one-dimensional FPU lattices.

On the other hand,
the existence (resp. non existence)
of small amplitude breathers
with frequencies slightly above the phonon band
has been established
when $V$ satisfies (resp. violates)
a local hardening condition~\cite{J01}.
The proof is based on a centre manifold technique,
which shows more generally that all small amplitude
time-periodic solutions of the FPU system (including breathers)
are determined by a two-dimensional map,
provided their frequency lies near the phonon band edge.
The discrete centre manifold method has been put in
a general framework~\cite{J03} and applied to other systems,
namely diatomic FPU chains~\cite{JN} (far from the uncoupled regime)
and spin lattices~\cite{N03}.

The aim of this work is to test
numerically the range of validity of the
centre manifold method~\cite{J01}
and to explore new phenomena, far from the small amplitude regime.

    The one-dimensional FPU system is given by the equations:
\begin{equation}
\label{eq:fpu} \ddot{x}_n = V'(x_{n+1}-x_n)-V'(x_n - x_{n-1})\; ,
\quad n\in \mathbb{Z} \;,
\end{equation}
where $x_n$ represents mass displacements from their equilibrium positions and
$V$ is a smooth interaction potential satisfying $V^{\prime}(0)=0$,
$V^{\prime\prime}(0)=1$. In ref.~\cite{J01}, the existence of small amplitude
breathers (SAB) with frequencies $\omega_{b}$ slightly above the phonon band
($\omega_{b} > 2$) is obtained for $B>0$, where
\begin{equation}
\label{defB}
B=\frac{1}{2}
V^{(4)}(0)-(V^{(3)}(0))^{2},
\end{equation}
and their non-existence is proved for $B<0$
(see \cite{T72}, \cite{F96} for related results
on the modulational instability of nonlinear
normal modes
and the tangent bifurcation of standing waves respectively).

The parameter $B$ can be interpreted as a hardening coefficient,
since breathers with amplitude $A\approx 0$ have frequency
$\omega_b \approx 2+ \frac{B}{8}\, A^2$.
Note that $B$ is slightly different
from the classical hardening coefficient of
an anharmonic potential
($V$ is hard if $\frac{3}{5}
V^{(4)}(0)-(V^{(3)}(0))^{2}>0$ and soft for
$\frac{3}{5}
V^{(4)}(0)-(V^{(3)}(0))^{2}<0$).

In this paper we consider anharmonic potentials
\be
V(u)=\frac{u^2}{2}+\frac{K_3}{3} u^3+\frac{K_4}{4} u^4
\quad
\label{eq:V}
\ee
(at least one of the coefficients $K_3$, $K_4$ is nonzero).
One has $B=3K_4-4 K_3^2$ and thus the parameter $B$ is positive if
$|K_3|<\sqrt{3 K_4}/2$. In the case $K_3=0$ we have an even potential and the
sign of $B$ coincides with the sign of the quartic coefficient $K_4$.

In section II we compute numerically SAB and find that they are very well
fitted by leading order analytical expressions deduced from ref.~\cite{J01}. In
section III we consider a class of potentials for which numerical computations
yield large amplitude breathers (LAB) with frequencies near the top of the
phonon band. In this case there is an energy threshold for the existence of
exact discrete breathers.
The existence of an energy threshold for breather creation in
higher-dimensional lattices is well known~\cite{FKM,W99},
but only a few
one-dimensional examples have been given
where such a threshold exists
\cite{MW,AKK01,Bernardo}.
Section III explores in more details the example of ref.
\cite{Bernardo}.
Section IV deals with dark breathers, i.e.
spatially modulated standing waves whose amplitude is constant at infinity
and vanishes at the center of the chain.
Leading order analytical
expressions for small amplitude dark breathers (with frequencies inside the
phonon band) fit the numerically computed dark breathers very well in the
case of even potentials. In section V we analyze bright and dark breathers
superposed on uniformly stretched or compressed static states. In each section
the linear stability of the above mentioned solutions is numerically
investigated.

\section{Small amplitude breathers}
\subsection{Summary of the theory}
We consider time-periodic solutions of (\ref{eq:fpu})
having a given frequency $\omega_b$.
We introduce the interaction force
\begin{equation}
y_n=V'(u_n)=u_n+K_3u_n^2+K_4u_n^3 \quad ,
\end{equation}
where \be u_n=x_n-x_{n-1}. \label{eq:un} \ee The time-average of $y_n$ is
independent of $n$ (integrate (\ref{eq:fpu})), and is fixed to $0$ in reference
\cite{J01}, as it is the case for all spatially localized solutions. Then,
problem (\ref{eq:fpu}) leads to
\begin{equation}
\label{dfpu}
\frac{d^{2}}{dt^{2}}{(V^{\prime})}^{-1}(y_{n})=y_{n+1}-2y_{n}+y_{n-1},
\ \ \ n\in \Z .
\end{equation}
The main results of~\cite{J01} can be formulated as follows.
For $\omega_b\approx 2$ (maximal phonon frequency),
all small amplitude solutions $y_n$ (even in $t$) have the form
\begin{equation}
\label{solexact}
y_{n}(t)=
b_{n}\cos{(\omega_b\, t)}+\varphi (t,b_{n-1},b_{n},\omega_b ),
\end{equation}
where $\varphi $ is a smooth time-periodic function with frequency $\omega_b$
and
$
\bsup{-1}{t\in [0,2\pi /\omega_b ]}
|\, \varphi (t,b_{n-1},b_{n},\omega_b )\, |
=
O(\, ( |b_{n-1}|+|b_{n}| )\, ( |b_{n-1}|+|b_{n}|+|\omega_b -2|^2 ) \, )$.
This result can be seen as an exact version of the
rotating-wave approximation method~\cite{ST88}, in
the small amplitude limit.
Note that $\varphi $ can be computed at an arbitrary order (see
\cite{J03} p. 51).
Moreover, $b_{n}$ satisfies the second order nonlinear
recurrence relation
\begin{equation}
\label{recrel}
b_{n+1}+2b_{n}+b_{n-1}=
-\mu\,b_{n}+B\, b_{n}^{3}+
\mbox{h.o.t}
\end{equation}
where $\mu=\wb^2-4\ll 1$ and
the coefficient $B$ is given in (\ref{defB}).
Higher order terms are
$O(\, |b_{n}|\,
{(  {(|b_{n-1}|+ |b_{n}|)}^{2}+|\mu |)}^{2}   )$
(note that the right side of (\ref{recrel}) can be computed
at an arbitrary order).
Consequently, the problem of finding small amplitude
time-periodic solutions of (\ref{dfpu}) reduces to
the problem (\ref{recrel}), which can be viewed
as a two-dimensional map
$(b_n,b_{n-1})\rightarrow (b_{n+1},b_{n})$.

For $B>0$ fixed and $\mu >0$ sufficiently small,
the recurrence relation (\ref{recrel}) has homoclinic solutions to $0$
satisfying $\lim\limits_{n\rightarrow\pm\infty}{b_n}=0$
(a proof has been given in \cite{J01}, \cite{J03},
using the invariance $n\rightarrow -n$ of (\ref{recrel})).
More precisely, there exist
homoclinic solutions to $0$ denoted as
$\pm\, b^1_n$, $\pm\, b^2_n$,
having different symmetries $b^1_{-n-1}=-b^1_n$, $b^2_{-n}=b^2_n$.

Beyond this particular result, one can expect the
existence of infinitely many homoclinic solutions to $0$
(not necessarily symmetric). Indeed, for $B>0$
the intersections of
the stable and unstable manifolds of $b_n=0$
are generically transverse, which yields
the existence of ``homoclinic tangles'' \cite{guck}.

A formal way to understand why homoclinic orbits exist is
the following. Setting $\xi_n =(-1)^n\, b_n$ yields the
recurrence relation
\be
\xi_{n+1}+ \xi_{n-1}-2\xi_n=\mu \xi_n-B\xi_n^3+ \mbox{h.o.t.}
\label{eq:map}\ee
Since $0<\mu\ll 1$, equation (\ref{eq:map}) can be approximated by an
integrable differential equation \be v''=v-v^3 \label{eq:v}  \ee with
\begin{equation}
\label{appsol}
\xi_n=\sqrt{\frac{\mu}{B}}\;\; v(n\sqrt{\mu}).
\end{equation}
Equation (\ref{eq:v}) has the homoclinic solutions
$v(x)=\pm \sqrt{2}/\cosh{(x+c)}$.

Using the above analysis, it has been proved
that for $B>0$ and
$\wb-2> 0$ sufficiently small, there exist SAB
solutions of (\ref{dfpu}) given by equation
(\ref{solexact}) and homoclinic solutions of (\ref{recrel})
(see \cite{J01}, \cite{J03}). SAB
have the form
\be y_n = (-1)^n \xi_n \cos(\omega_b t)+ \mbox{h.o.t.} \quad , \label{eq:y}\ee
where $\xi_n =(-1)^n\, b_n$ satisfies the recurrence relation
(\ref{eq:map})
and $|\xi_n|\rightarrow 0$ as $n\rightarrow \pm \infty$.
The above homoclinic solutions $b^1_n$, $b^2_n$ correspond
via equation (\ref{solexact})
to SAB $y^1_n$, $y^2_n$ having different symmetries
$y^1_{-n-1}(t)=y^1_n(t+T_b /2)$, $y^2_{-n}=y^2_n$,
where $T_b=2\pi /\omega_b$ is the breather period
(see \cite{J03} p.58,59).
Note that $y_n(t+T_b /2)\neq -y_n(t)$ in general
due to higher order terms in (\ref{eq:y})
(however the egality holds if $V$ is even).
Homoclinic solutions $-b^i_n$ of (\ref{recrel})
simply correspond to $y^i_n(t+T_b /2)$.

The exact solutions $y^1_n$, $y^2_n$ can be approximated
at leading order using (\ref{eq:v})-(\ref{appsol})
with $v(x)=\sqrt{2}/\cosh{x}$.
This yields
\be
y^2_n(t)\simeq (-1)^n \sqrt{\frac{2\mu}{B}} \frac{\cos \wb t}{\cosh(n\sqrt{\mu})}
\label{eq:SAB2} \quad ,
\ee
and by symmetry we construct the second approximation
\be
y^1_n(t)\simeq (-1)^n \sqrt{\frac{2\mu}{B}} \frac{\cos \wb t}{\cosh(\,
(|n+1/2|-1/2)\, \sqrt{\mu}\, )} \label{eq:SAB}  . \ee Under these
approximations one has $y^i_n(t+T_b /2)\approx -y^i_n(t)$ and $y^1_0(t)\approx
y^2_0(t)$. In fact, the exact solutions satisfy $y^i_n(t+T_b
/2)=-y^i_n(t)+O(|\mu |)$ and $y^1_0(t)= y^2_0(t)+O(|\mu |^{3/2})$ as $\mu
\rightarrow 0^+$.

Notice that approximation (\ref{eq:SAB2})
can also be derived
using multiscale expansions~\cite{T72}. Moreover, these
expressions also
approximate $u_n$ since $u_n =y_n +O(y_n^2)$.

These calculations show that the
maximum amplitude
$A\approx \sqrt{\frac{2\mu}{B}}$
of the SAB is
$O(|\omega_b -2|^{1/2})$ as $w_b \rightarrow 2^{+}$.
More precisely, breathers with amplitude $A\approx 0$ have frequency
$\omega_b \approx 2+ \frac{B}{8}\, A^2$.

Although SAB decay
exponentially as $|n|\rightarrow +\infty$, their width
is $O(|\omega_b -2|^{-1/2})$ and
diverges as $w_b
\rightarrow 2^{+}$. As a
consequence, if $B>0$ breathers exist for any small value of energy in the FPU
system (\ref{eq:fpu}).

\subsection{Numerical continuation}
    We have computed these solutions
numerically in order to check the range of validity of
approximations (\ref{eq:SAB2}), (\ref{eq:SAB}).
We have fixed $K_4=1$ in eq. (\ref{eq:V}) and thus the parameter $B$ is
positive if $|K_3|<K_3^*=\sqrt{3}/2\simeq 0.86$.

We have performed our computations using a numerical scheme based
on the concept of anti-continuous limit and path continuation with the Newton
method~\cite{MA}. With this technique it is more convenient to use the
difference displacements variables $u_n$.
Indeed, with these new variables the dynamical equations become
\begin{equation}
 \label{eq:dynu}
\ddot{u}_n+2V'(u_n)-C\left[V'(u_{n+1})+V'(u_{n-1})\right]=0 \; ,\quad  n\in
\mathbb{Z} \;,
\end{equation}
where $C=1$ for our system, but it makes possible to consider the system
(\ref{eq:dynu}) as a Klein--Gordon system, with an on--site potential $V$ and
an anticontinuous limit at $C=0$. Note that $u_n$ is one-to-one related to the
forces $y_n$ at small amplitudes, because $V'$ is locally invertible since
$V^{\prime\prime}(0)\neq 0$.

    We  use periodic boundary conditions $u_{n+2p}(t)=u_{n}(t)$.
The periodicity is considered for an even number of sites in order that the
maximum frequency of the linear phonons is exactly 2 as in the infinite lattice
(this frequency corresponds to the phonon with wave number $\pi$).

We have obtained breathers with symmetries $u_n(t)=u_{-n}(t)$ (site-centred
mode) and $u_n(t)=u_{-n-1} (t+T_{b}/2)$ (bond-centred mode).
Note that the coordinates
transformation (\ref{eq:un}) produces an exchange of the symmetry properties
between both modes, i.e. in the difference displacements variables the
site-centred mode is the Page mode~\cite{Pag90} and the bond-centred mode is
the Sievers-Takeno mode~\cite{ST88}.

\begin{figure}
\begin{center}
\includegraphics[width=8cm]{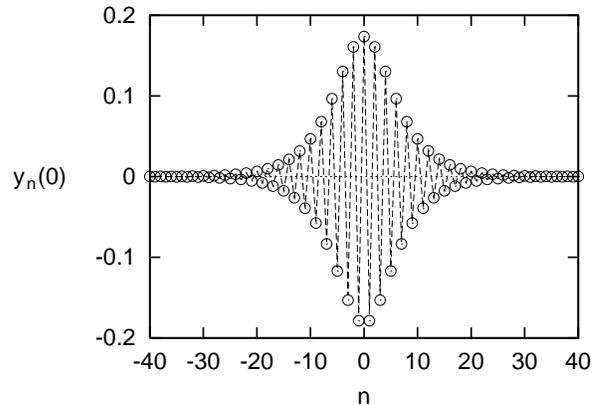}
\end{center}
\caption{Comparison between an exact site-centred mode (circles) and the
approximation (\ref{eq:SAB2}) (dashed line) for $\wb=2.01$ ($\mu\approx 0.04$). The cubic coefficient of the potential is $K_3=-0.3$ ($B=2.64$),
but similar results are obtained for any other fixed value of $K_3$
with $|K_3 |<K_3^*$ ($B>0$) and $\mu >0 $ small enough.
\label{fig1}}
\end{figure}

     Fig.\ref{fig1} shows the excellent agreement between the approximation
(\ref{eq:SAB2}) (dashed line) and an exact site-centred mode obtained
numerically (circles) for $\wb=2.01$ ($\mu \approx 0.04$).

\begin{figure}
\begin{center}
\includegraphics[width=8cm]{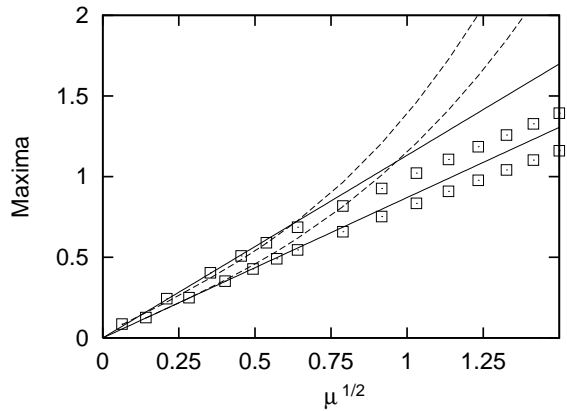}
\end{center}
\caption{Force (dashed line) and relative displacements
(squares) maxima versus $\mu^{1/2}$.
The continuous line corresponds to the maxima given by approximation
(\ref{eq:SAB2}). Parameters from top to bottom: $K_3=-0.6$ ($B=1.56$) and
$K_3=-0.3$ ($B=2.64$). \label{fig2}}
\end{figure}

    We have continued the SAB obtained numerically for $B>0$ as $\wb$ goes away
from the phonon band. In Fig.\ref{fig2} we have plotted the maximum amplitude
of the forces sup$|y_n|$ (dash lines) versus $\mu^{1/2}$ for two different
values of the cubic coefficient, $K_3=-0.6$ ($B=1.56$) and $K_3=-0.3$
($B=2.64$) from top to bottom, respectively.  The maximum amplitude of the
forces is an approximately linear function of $\mu^{1/2}$ up to $\mu \approx
0.6$. The continuous lines represent the maxima given by approximation
(\ref{eq:SAB2}).
This approximation is not satisfactory for very small values of $B$ since
higher order terms in (\ref{recrel}) cannot be neglected when $B\rightarrow
0$. More precisely, the smaller $B$ is, the smaller $\mu$ must be chosen
in (\ref{eq:SAB2}) in order to have a good approximation.

\begin{figure}
\begin{center}
\includegraphics[width=8cm]{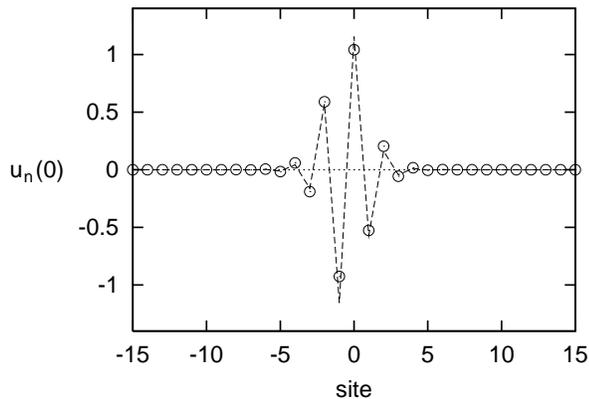}
\end{center}
\caption{Comparison between an exact bond-centered mode (circles) and the
analytical approximation (\ref{eq:SAB}) (dashed line) for $\mu=1.76$.  The
value of $\mu$ corresponds to a breather frequency $\wb=2.4$ relatively far
from the phonon band. The cubic coefficient of the potential $V$ is $K_3=-0.3$
($B=2.64$).  \label{fig3}}
\end{figure}

    The maximum amplitude of the relative displacements sup$|u_n|$  is also an
approximately linear function of $\mu^{1/2}$ (see squares in fig.\ref{fig2}).
This was expected for small $\mu$ since $u_{n}=y_{n}+O(y_{n}^{2})$, but it
occurs far from the phonon band, at least until values of $\mu \approx 2.25$ .
In fact, we have checked that, surprisingly, expressions (\ref{eq:SAB2}) and
(\ref{eq:SAB}) fit very well the profile of the relative displacements $u_n$
even far from the phonon band as Fig. \ref{fig3} shows. Note that the vibration
amplitudes of the breather in Fig. \ref{fig3} are quite large.

\begin{figure}
\begin{center}
\includegraphics[width=8cm]{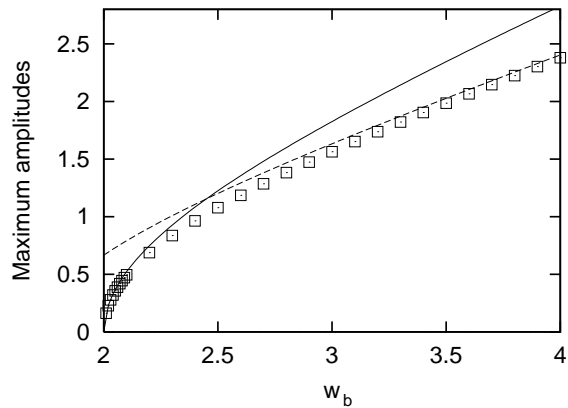}
\end{center}
\caption{Maximum amplitudes, $sup|u_n|$, of the Sievers-Takeno mode versus
frequency for an even potential.  The continuous line is the maxima given by
eq. (\ref{eq:maxima}), while the dashed line corresponds to the
Sievers-Takeno approximation.
Squares correspond to the numerically computed solution.
\label{fig4}}
\end{figure}

    In Fig~\ref{fig4} we have plotted again the maximum amplitude
of the relative displacements
(squares), but now for an even potential ($K_3=0$) and larger frequencies,
in order to compare approximation (\ref{eq:SAB})
with an approximation derived by Sievers and Takeno
in the large amplitude regime \cite{ST88}.
The Sievers-Takeno approximation is obtained using the
rotating wave approximation (only the first Fourier component of
solutions is taken into account) and Green's function techniques.
For the Sievers-Takeno mode and $K_4=1$ ($B=3$) it predicts
\begin{equation}
\label{sta}
sup|u_n|\approx \frac{2}{3} \sqrt{w_b^2-3} .
\end{equation}
This amplitude-frequency relation corresponds to the dashed line in
Fig~\ref{fig4}. On the other hand, from eq. (\ref{eq:SAB}) it follows
(continuous line)
\begin{equation} sup|u_n|\approx \sqrt{\frac{2(w_b^2-4)}{3}} .
\label{eq:maxima} \end{equation}
Clearly we see that approximation (\ref{sta})
fits better very localized solutions with large
amplitudes, whereas approximation
(\ref{eq:maxima}) works better up to moderate amplitudes
($\omega_b \approx 2.4$ in this case).
In this sense both approximations can be considered complementary.

\begin{figure}
\begin{tabular}{c}
\includegraphics[width=8cm]{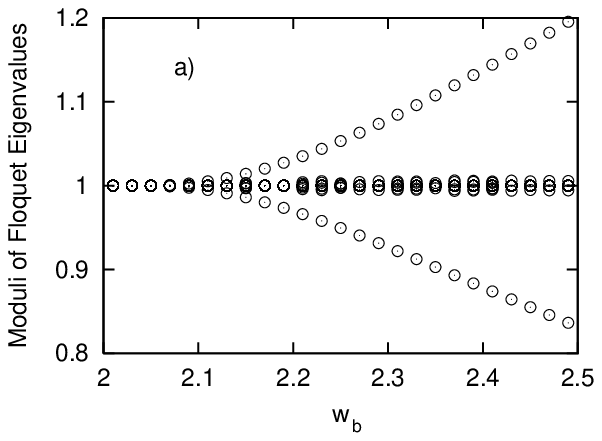}\\
\includegraphics[width=8cm]{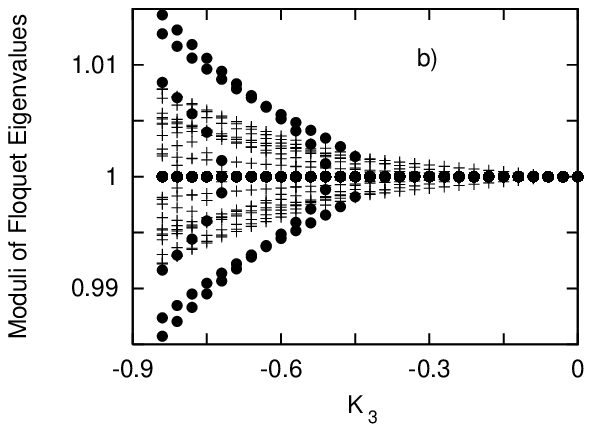}
\end{tabular}
\caption{a) Harmonic instability of the Sievers-Takeno mode for $K_3=-0.3$; b)
Oscillatory instabilities of the Page mode versus $K_3$ for a lattice with 52
sites (full circles) and a lattice with 100 sites (pluses). The frequency is
$w_b=2.1$. \label{fig5}}
\end{figure}

    In addition we have performed a Floquet analysis in
order to study the linear stability of the SAB~\cite{Aubry}. In the symmetric
potential case the result is well known~\cite{Bickham}: the Page mode is stable
while the Sievers-Takeno mode has a harmonic instability (a pair of real
eigenvalues $\sigma ,\sigma^{-1}$ close to $1$) that increases with the
breather frequency. When a cubic term in the potential $V$ is introduced
$(K_3\ne 0)$, the situation is more complex. The Sievers-Takeno mode shows
again a harmonic instability (see Fig~\ref{fig5}a) but it also appears
oscillatory instabilities. As Fig~\ref{fig5}b shows for the Page mode, these
oscillatory instabilities increase with $|K_3|$.  Although most of them are
size dependent effects due to the discretization of continuous
spectrum~\cite{MA98}, for the infinite system
some of them might remain out of the unit circle
and in this case SAB would be unstable.
We shall not examine this problem here
and leave it for future works.

\section{Large amplitude breathers with an energy threshold}
In this section we study breathers in the parameter region $B<0$.
Using the
recurrence relation (\ref{recrel}), one can show that SAB with frequencies
slightly above the phonon band do not exist for $B<0$ (see \cite{J01},
\cite{J03}). However, large amplitude breathers (LAB) do exist in this
parameter region \cite{AKK01}. More precisely, for $V$ strictly convex
($K_4>0$, $|K_3|<\sqrt{3K_4}$) and  $B<0$ ($\sqrt{3K_4}/2<|K_3|$),
breathers whose amplitudes do not tend to zero as $w_b \rightarrow 2^{+}$
have been numerically obtained \cite{AKK01,Bernardo,Kastner}.
As a
consequence there is an energy threshold for breathers creation in these FPU
systems.
The criterion $\sqrt{3K_4}/2<|K_3|<\sqrt{3K_4}$
for the existence of an energy threshold originates from reference
\cite{Bernardo}, and follows by combining analytical results in
\cite{J03} (non existence of small amplitude breathers) and
\cite{AKK01} (existence of large amplitude breathers).

The existence of an energy threshold for breather creation
has been formally analyzed by Flach, Kladko and MacKay
\cite{FKM} for general Hamiltonian systems,
in a different case when small amplitude breathers bifurcate
at an edge of the phonon spectrum.
An energy threshold exists above a critical lattice dimension $d$
(typically $d=2$) which depends on the nonlinear terms of
the Hamiltonian. This analysis has been extended to a class
of non-analytic Hamiltonian systems in \cite{Kastner}.
The existence of
excitation thresholds has been rigorously analyzed by
Weinstein \cite{W99} for the $d$-dimensional
discrete non linear Schr\"odinger
equation, using variational methods
(see also \cite{MW} for formal variational arguments in
the one-dimensional case).

\begin{figure}
\begin{center}
\includegraphics[width=8cm]{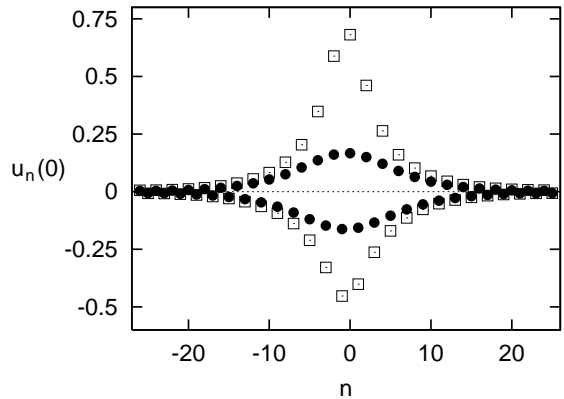}
\caption{Comparison between a small amplitude breather (full circles) for $B>0$
and a large amplitude breather (blank squares) for $B<0$ with the same
frequency $w_b=2.01$. Parameters:  SAB: $K_3=-0.3$, $K_4=1$ ($B=2.64$); LAB:
$K_3=-1$, $K_4=1$ ($B=-1$).   \label{fig6}}
\end{center}
\end{figure}

\begin{figure}
\begin{tabular}{c}
\includegraphics[width=8cm]{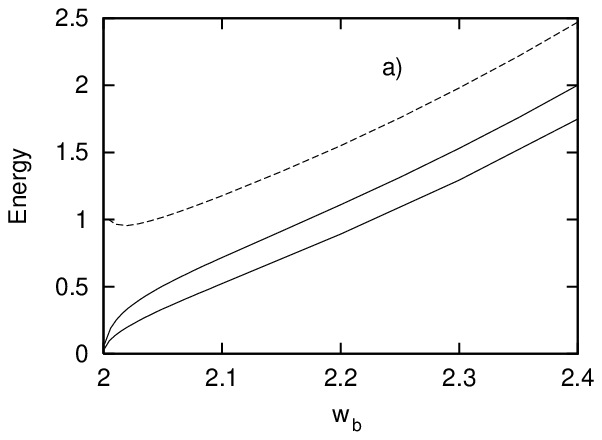}\\
\includegraphics[width=8cm]{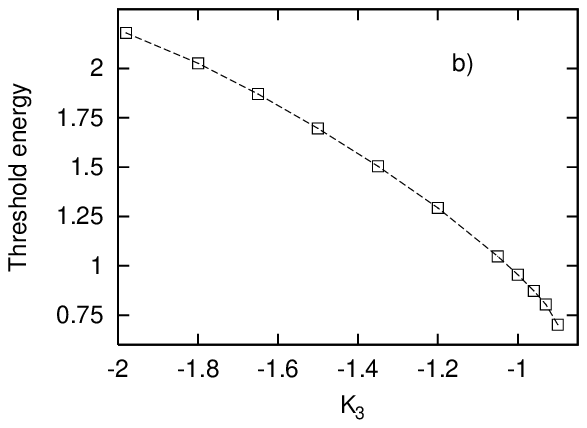}
\end{tabular}
\caption{(a) Breather energy versus frequency close to the phonon band edge (we fix $K_4 =1$).
The dashed line shows the existence of a threshold in energy for a LAB with
$K_3=-1$ ($B<0$). The continuous lines ($K_3=-0.6$ and $K_3=0$ from top to bottom) show
that for $B>0$ SAB of arbitrary low energy exist. (b) Dependence of the LAB
energy threshold on the asymmetry parameter $K_3$. \label{fig7}}
\end{figure}

    In Fig. \ref{fig6} we compare a SAB and a LAB with the same frequency $w_b=2.01$.
As it is shown in Fig~\ref{fig7}a the energy of a LAB (dash line) does not go
to zero when approaching the phonon band edge. It lies always above a certain
positive lower bound. In the case of SAB (continuous lines) the breather energy
can be lowered to arbitrarily small values.

The dependence of the energy
threshold on the asymmetry parameter $K_3$ is also shown in Fig.~\ref{fig7}b,
where we have fixed $K_4 =1$.
One observes that
the energy threshold goes to $0$ as $B\rightarrow 0^-$
(for $K_3\rightarrow -\sqrt{3}/2\approx -0,86$).
Moreover, LAB families
also exist for nonconvex potentials
($K_3<-\sqrt{3}\approx -1,73$), and
their energy threshold increases with
$|K_3|$. This shows that the
criterion $\sqrt{3K_4}/2<|K_3|<\sqrt{3K_4}$ for
the existence of an energy threshold is not optimal.
We conjecturate that an energy threshold appears
under the more general condition $K_4 >0$, $\sqrt{3K_4}/2<|K_3|$
($B<0$) in the one-dimensional FPU system.

\begin{figure}
\begin{center}
\includegraphics[width=8cm]{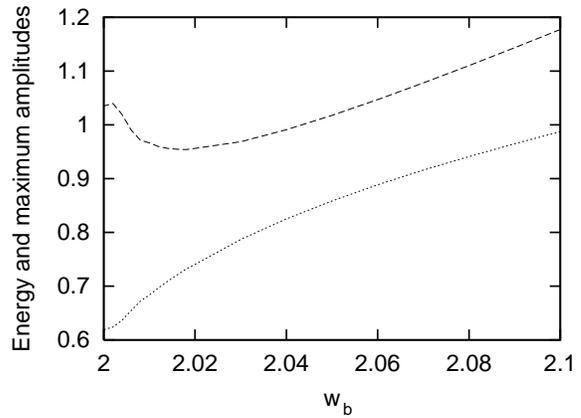}
\end{center}
\caption{ LAB energy (upper line) and amplitude (lower line) vs frequency close
to the phonon band edge (we fix $K_3=-1$, $K_4 =1$). \label{fig8}}
\end{figure}

Figure \ref{fig8} shows the energy of a LAB (upper line) and its amplitude
(lower line) versus frequency, close to the phonon band edge (the upper line is
a close-up of figure \ref{fig7}a near $\omega_b =2$). The reason for the non
monotonous behaviour of energy is that there are two competing tendencies: on
the one hand amplitudes are decreasing with frequency and going to a nonzero
lower bound, but on the other hand the width of the breather is increasing and
more and more particles begin to oscillate as $\omega_b$ decreases.

\begin{figure}
\begin{center}
\includegraphics[width=8cm]{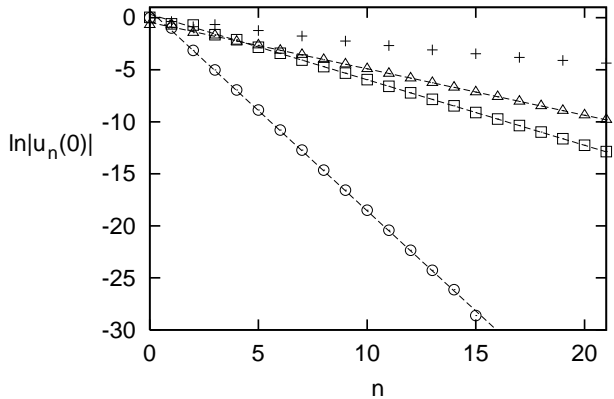}
\end{center}
\caption{Logarithms of the Page mode profiles for $K_3=-1$, $K_4=1$ and $w_b=3$
(circles),  $w_b=2.1$ (squares), $w_b=2.05$ (triangles) and $w_b=2$ (pluses) .
The dashed line is the logarithm of eq.(\ref{eq:lab}) after fitting the
parameter $\alpha$. Note that  eq.(\ref{eq:lab}) does not work in the limit
 $\omega_b=2$. \label{fig9}}
\end{figure}

We have found LAB with the same symmetries as SAB (Page and Sievers-Takeno
modes). LAB have an exponential decay for $\omega_b >2$. As Fig. \ref{fig9}
shows,  we have been able to fit the LAB profiles to the following expression
in the case of the Page-mode: \be \label{eq:lab} u_n(t)\approx \alpha
(\omega_b)\, \sigma (\omega_b)^{|n|} \cos{(\omega_{b} t)}
 \ee
  where
  \be \sigma
(\omega_b )= 1-\frac{\omega_b^2}{2}+ \frac{\omega_b}{2} (\omega_b^2-4)^{1/2}
\in (-1,0) \quad ,
 \ee
and $\alpha (\omega_b)$ is a fitting parameter. A similar expression
corresponds to the Sievers-Takeno mode.

We have computed the decay rate $|\sigma (\omega_b) |$ as follows. The FPU
system (\ref{eq:fpu}) can be formulated as a mapping \be \binom{y_n}{y_{n+1}}
=F \binom{y_{n-1}}{y_{n}}
 \ee where the map
\be F \binom{y_{n-1}}{y_{n}}= \binom{y_n}{\frac{d^2}{dt^2}
(V')^{-1}(y_n)+2y_n-y_{n-1}} \ee acts in a space of smooth time-periodic
functions with frequency $\omega_{b}>2$. The linearized operator \be
DF(0)\binom{y_{n-1}}{y_{n}}= \binom{y_n}{(\frac{d^2}{dt^2} +2)y_n-y_{n-1}} \ee
has a purely hyperbolic spectrum and $\sigma(\omega_b )$,
$\sigma(\omega_b)^{-1}$ are the closest eigenvalues to $-1$ (with
$\sigma(2)=-1$).

We have observed that LAB remain localized for $\omega_b=2$ (see pluses in
figure \ref{fig9}) and thus approximation (\ref{eq:lab}) does not work in this
limit.

\begin{figure}
\begin{center}
\includegraphics[width=8cm]{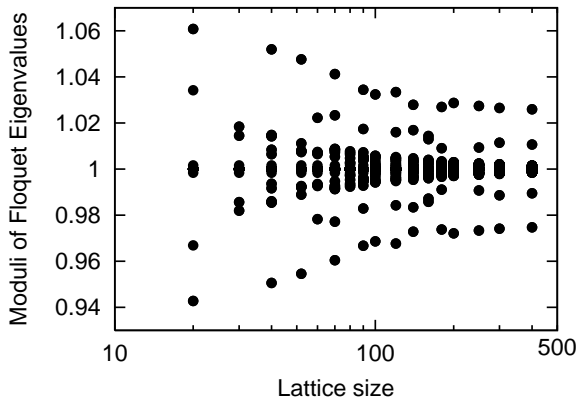}
\end{center}
\caption{ Floquet eigenvalues of the Page mode for $\omega_b =3$, $K_3=-1$ and
$K_4=1$, depending on system size. The eigenvalue of largest modulus and its
inverse seem to converge to values out of the unit circle for large system
sizes. \label{fig10}}
\end{figure}

    The stability properties of LAB are the same as SAB with non-even
potentials, i.e. both modes show oscillatory instabilities and also a harmonic
instability in the case of the Sievers-Takeno mode. As we mention in section
II, most of the oscillatory instabilities are size dependent effects due to the
discretization of continuous spectrum \cite{MA98}. However, a band analysis
\cite{AKK01} suggests that, at least for some parameter values (close to
$\omega_b =3$, $K_3=-1$, $K_4=1$), some eigenvalues remain out of the unit
circle for the infinite system and LAB are unstable. We have studied the
evolution of Floquet eigenvalues of the Page mode for $\omega_b =3$, $K_3=-1$
and $K_4=1$ when the system size becomes larger. In figure \ref{fig10} one can
see that one pair of eigenvalues seems to converge to values out of the unit
circle for large system sizes, as stated in \cite{AKK01}.

\section{Dark Breathers}
Using the recurrence relation (\ref{recrel}),
it has been shown
that for $B<0$ and $\mu<0$ small enough
($\omega_b$ lies in the phonon band)
there exist small amplitude
heteroclinic solutions of (\ref{dfpu})
connecting two nonlinear normal modes
$y^{\pm}_n$ at infinity,
i.e.
$\lim\limits_{n\rightarrow \pm \infty}{y_{n}(t)-y_n^\pm (t)}=0$
(see \cite{J01}, \cite{J03}).

We now briefly describe the nonlinear normal modes $y^{\pm}_n$.
These solutions are spatially periodic
($y^{\pm}_{n+2}(t)=y^{\pm}_n(t)$), have $0$ time-average
(the mean interaction force is $0$) and neighbours oscillate out of phase
($y^{\pm}_{n+1}(t)=y^{\pm}_n(t+T_b /2)$).
Moreover, solutions $y^{\pm}_n$ are equal up to a half period phase shift
($y^{+}_{n}(t)=y^{-}_n(t+T_b /2)$).
Note that the corresponding displacement patterns $x_n$
are in general different from the zone-boundary modes
$x_n(t)=(-1)^n\, f(t)$ (also called binary oscillations). Indeed,
in this latter case the time-average of relative displacements $u_n$
is $0$, whereas interaction forces have $0$ time-average
in our case.
However, both types of solutions coincide for even potentials $V$.

To obtain
an approximate expression of the heteroclinic solutions,
we can rewrite the recurrence relation
(\ref{eq:map}) in the following form
\be \frac{\rho_{n+1}+ \rho_{n-1}-2\rho_n}{-\mu}=-\rho_n-B\rho_n^3+ \mbox{h.o.t.}
\label{eq:map2} \quad ,\ee
using the variable change $\xi_n=\sqrt{-\mu} \; \rho_n$. Equation
(\ref{eq:map2}) can be approximated by the differential equation:
\be v''=-v-Bv^3 \ee
with $\rho_n=v(n\sqrt{\mu})$.
Consequently, the system (\ref{dfpu}) has a family of
small amplitude solutions which can be approximated by
\be y_n (t) \simeq (-1)^n \sqrt{\frac{\mu}{B}} \tanh\left(\frac{n\sqrt{-\mu}}{\sqrt{2}}\right)
\cos(\wb t)   .       \label{eq:dark}\ee
This type of solutions are called
dark breathers~\cite{dark} or antisymmetric kinks~\cite{BFW97}. Typically the
term breathers refers to bright breathers, that is a few excited units in
a lattice. Here we
have the opposite kind of localization: most oscillators are excited except
one or a few of them.

Expression (\ref{eq:dark}) approximates exact site-centred
solutions of (\ref{dfpu}) satisfying $y_{-n}(t)=y_{n}(t+T_b/2)$.
Under our approximation one has
$y_n(t+T_b /2)\approx -y_n(t)$, and
in fact the exact solutions satisfy
$y_n(t+T_b /2)=-y_n(t)+O(|\mu |)$ (uniformly in $n\in\Z$)
as $\mu \rightarrow 0^+$.

Equation (\ref{dfpu}) has also exact bond-centred
dark breather solutions satisfying $y_n(t)=y_{-n+1}(t)$.
We shall approximate them by
\be y_n (t) \simeq (-1)^n \sqrt{\frac{\mu}{B}}
\tanh\left(\frac{(-n+1/2)\sqrt{-\mu}}{\sqrt{2}}\right)
\cos(\wb t)   .       \label{eq:darkbc}\ee
Note that
expressions (\ref{eq:dark}), (\ref{eq:darkbc}) also approximate
$u_n$ since $u_n =y_n +O(y_n^2)$.

\begin{figure}
\includegraphics[width=8cm]{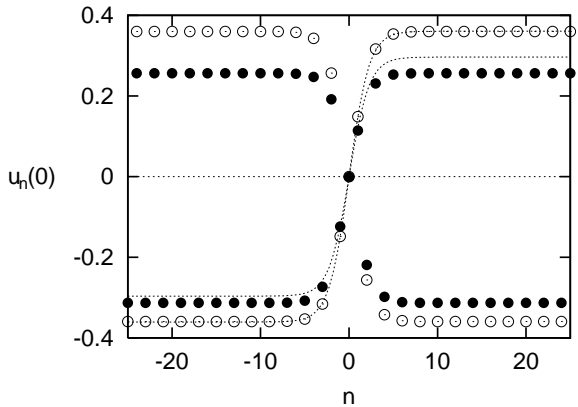}
\caption{Small amplitude site-centred dark breathers for a frequency $\wb=1.9$.
The circles represent an exact solution computed numerically with an even
potential ($K_4=-1$), while the full circles were obtained adding  to the
previous potential a cubic term with a coefficient $K_3=0.6$. These solutions
have the symmetry $u_{-n}(t)=u_{n}(t+T_b/2)$. The dashed lines are the $\tanh$
function of approximation (\ref{eq:dark}) for both cases. \label{fig11}}
\end{figure}

\begin{figure}
\includegraphics[width=8cm]{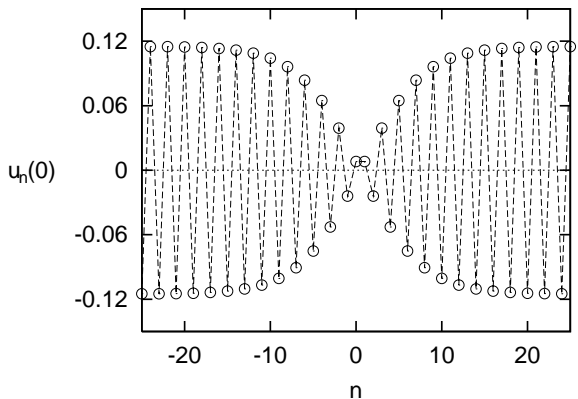}
\caption{ Small amplitude bond-centred dark breather for a even potential with
$K_4=-1$ and a frequency $\wb=1.99$. The circles represent an exact solution
computed numerically, having the symmetry $u_n(t)=u_{-n+1}(t)$. The dashed line
corresponds to the analytical approximation (\ref{eq:darkbc}). \label{fig12}}
\end{figure}

For computing dark breathers numerically we consider a lattice with an
odd number of sites ($51$ lattice sites).
We first choose an even potential ($K_3=0$) with $K_4 <0$
(so that $B<0$).
We solve equation (\ref{eq:dynu}) in Fourier space by the Newton method,
fixing the time-average of $u_0$ to $0$ and
using (\ref{eq:dark}) as an initial condition.
Dark breathers are computed for non-even potentials by
continuation with respect to $K_3$, starting from $K_3=0$
(during the continuation the time-average of $y_n$
remains very small, due to the constraint on $u_0$).

Fig.~\ref{fig11} shows the excellent agreement between an exact site-centred
solution obtained numerically (circles) and the approximation (\ref{eq:dark})
(dashed line), for  an even potential with $K_4=-1$ ($B=-3$) and a frequency
$\wb=1.9$. For non even potentials (full circles) the agreement is not so good.
The reason is clearly that the approximation of $u_n$ by $y_n$ with equation
(\ref{eq:dark}) has the symmetry property $u_{n}(t+T_{b}/2)=-u_n(t)$ which is
not true any more when $K_3\ne0$.

Similar results are obtained for bond-centred solutions (see Fig.~\ref{fig12}).

\begin{figure}
\begin{center}
\includegraphics[width=8cm]{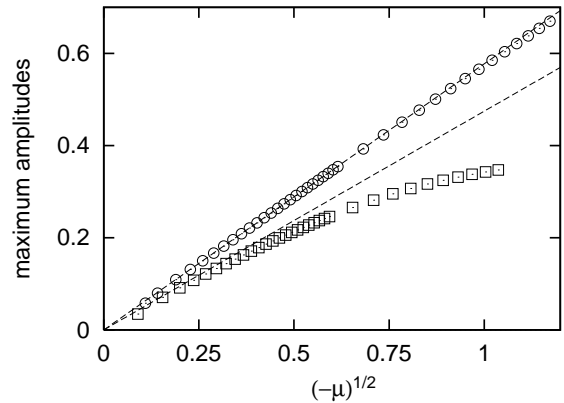}
\end{center}
\caption{Maximum amplitudes, $sup|u_n|$, of the site-centred dark breather
versus $-\mu^{1/2}$ for $K_3=0$ (circles) and $K_3=0.6$ (squares). In both
cases $K_4=-1$. The dashed line is the maxima given by eq. (\ref{eq:dark}). In
the non-even case approximation (\ref{eq:dark}) does not provide a good fitting
far from the phonon band edge. \label{fig13}}
\end{figure}

For a fixed potential we can continue numerically these  solutions decreasing
$\wb$. As Fig.~\ref{fig13} shows,  then the oscillation amplitudes increase and
for even potentials  expression (\ref{eq:dark}) continues matching very well
the relative displacements $u_n$ (circles), even far from the top of the phonon
band. For non even potentials (squares) the agreement is also excellent close
enough to the edge of the phonon band. For lower frequencies we observe
discrepancies becoming larger.

We have also performed a Floquet analysis to
investigate the linear stability
of these family of solutions when we decrease the frequency.

\begin{figure}
\begin{tabular}{c}
\includegraphics[width=8cm,height=4cm]{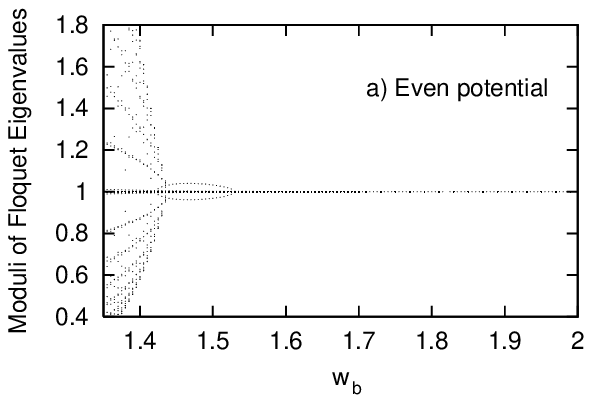}\\
\includegraphics[width=8cm,height=4cm]{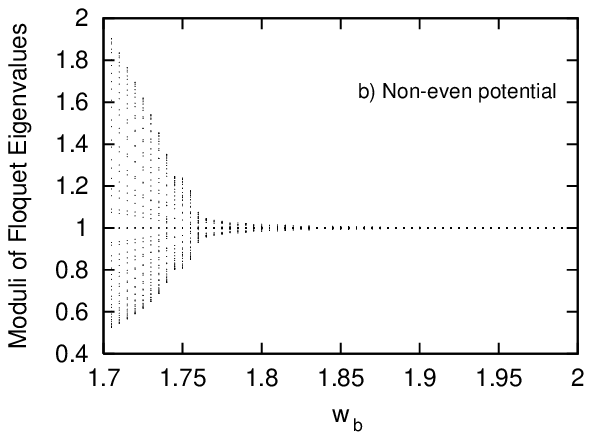}
\end{tabular}
\caption{Evolution of the Floquet eigenvalue moduli versus frequency for the
site-centred dark breather. a) Even potential ($K_3=0$); b) Non-even potential
($K_3=-0.6$). In both cases $K_4=-1$. \label{fig14}}
\end{figure}

    On the one hand, the site-centered mode is stable close to the edge of the phonon band for
even potentials as figure~\ref{fig14}a shows. However, a cascade of harmonic
instabilities appears for low enough frequencies (larger amplitudes). These
instabilities are visible in Fig.~\ref{fig14}a at $\wb \approx 1.43$. Before
coming out the harmonic instabilities one observes an instability bubble at
$\wb \approx 1.53$. Nevertheless, its magnitude decreases as the system size
increases and thus it should not be relevant for the infinite system.

\begin{figure}
\begin{tabular}{c}
\includegraphics[width=8cm,height=4cm]{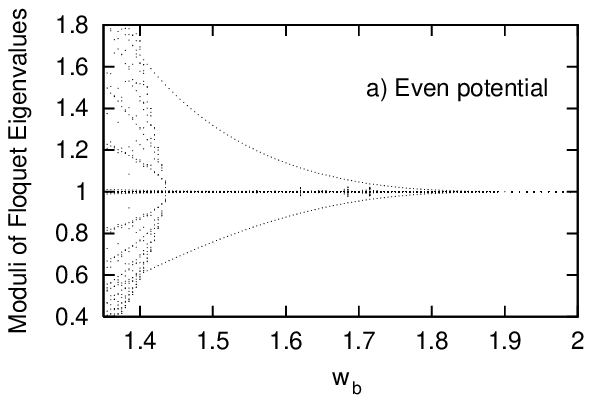}\\
\includegraphics[width=8cm,height=4cm]{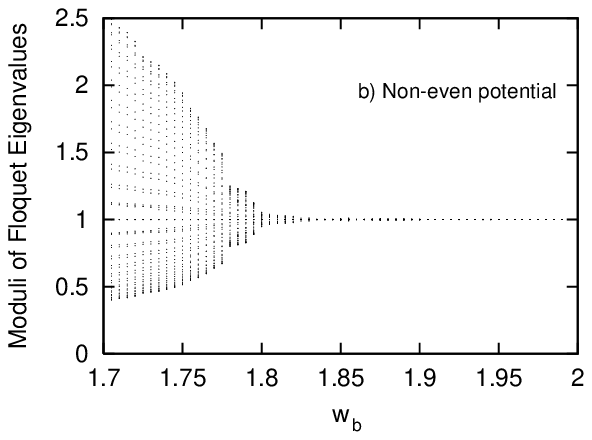}\\
\end{tabular}
\caption{Evolution of the Floquet eigenvalue moduli versus frequency for the
bond-centred dark breather. a) Even potential ($K_3=0$); b) Non-even potential
($K_3=-0.6$). In both cases $K_4=-1$.  \label{fig15}}
\end{figure}

    On the other hand, the bond-centered mode shows a harmonic instability (see
figure~\ref{fig15}a) that increases when the frequency decreases. One also
observes another cascade of harmonic instabilities  at $\wb \approx 1.43$.

Non-even potentials induce oscillatory instabilities for low enough frequencies
(larger amplitudes), both for the site-centered and the bond-centered modes.
Fig.~\ref{fig14}b and fig.~\ref{fig15}b show that this occurs for frequencies
lower than $1.85$, for $K_3=-0.6$ and $K_4=-1$. Numerically we haven't found
these oscillatory instabilities near the phonon band edge, but we do not
exclude that they might exist and be harder to detect.

The harmonic instability before mentioned for the bond-centered mode is also
present for non even potentials, but is not visible in fig.~\ref{fig15}b due to
oscillatory instabilities. Note that frequency scales in figures a) and b) are
different. For this reason in figures b) we do not see  either the cascade of
harmonic instabilities around $\wb \approx 1.43$. We have checked that they are
also present with non-even potentials.


\section{ Bright and dark breathers with an uniform stress \label{stress}}
\begin{figure}
\begin{center}
\includegraphics[width=8cm]{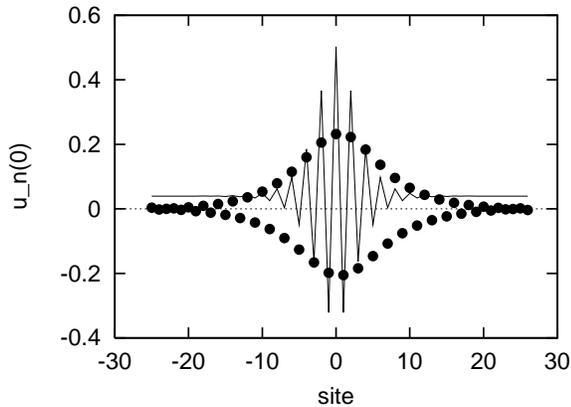}
\end{center}
\caption{Comparison between a SAB (full circles) and a breather with an uniform
stress (continuous line) for the same values of the parameters: $K_3=-0.6$,
$K_4=1$, $w_b=2.01$. \label{fig16}}
\end{figure}

    Breathers considered in section III for $B<0$ can be numerically continued at fixed
frequency up to values $B>0$, bringing about a new type of solutions which
satisfy $\lim_{n\rightarrow \pm \infty} u_n=c$, where $c$ is a nonzero
constant. As Fig.~\ref{fig16} shows, these new type of solutions can coexist
with SAB for the same values of the parameters and can be seen as time-periodic
oscillations around static solutions of (\ref{eq:fpu}) given by $x_n =c\, n$.
The lattice is uniformly stretched at infinity for $c>0$, and uniformly
compressed for $c<0$. Such breather solutions with a constant static strain
$c<0$ have been also observed by Sandusky and Page in the case $B<0$ (see
\cite{SanduskyP} p.876).

This section provides analytical and numerical results
concerning breather solutions with a constant static strain.
The case of dark breathers with an uniform stress
is also examined.

In the sequel we assume ${V}^{\prime\prime}(c)>0$.
We make the change of variable
$x_n (t)=c n+\tilde{x}_n(t\, \sqrt{{V}^{\prime\prime}(c)})$ where
$\tilde{x}_n \ll 1$ and introduce the
modified potential
\begin{equation}
\label{modpot}
\tilde{V}(u)=(V(c+u)-V^{\prime}(c)\, u)/{V}^{\prime\prime}(c).
\end{equation}
Equation (\ref{eq:fpu}) leads to
\begin{equation}
\label{eq:fpuc}
\frac{d^2}{dt^2}{\tilde{x}_n} =
\tilde{V}'(\tilde{x}_{n+1}-\tilde{x}_n)
-\tilde{V}'(\tilde{x}_n -\tilde{x}_{n-1})\; ,
\quad n\in \mathbb{Z} \;,
\end{equation}
with
$\tilde{V}^{\prime}(0)=0$ and
$\tilde{V}^{\prime\prime}(0)=1$.
One can see that equation (\ref{eq:fpuc}) is exactly
equation (\ref{eq:fpu}) where $V$ has been changed to
$\tilde{V}$. Consequently, the results of the above
sections (both analytical and numerical) readily
apply to equation (\ref{eq:fpuc}).

Returning to the original variables,
the following analytical results for the FPU system
(\ref{eq:fpu}) follow from the analysis of section II-A.
Let us fix $c\in \R$ and define
\begin{equation}
\label{defBc}
B(c)=\frac{1}{2}
{V}^{\prime\prime}(c)\,
V^{(4)}(c)-(V^{(3)}(c))^{2}.
\end{equation}
If $B(c)>0$ and
$\omega_b - 2\sqrt{{V}^{\prime\prime}(c)}>0$
is sufficiently small,
the FPU system has a family of exact
solutions
which can be approximated by
\be
u_n(t)\simeq
c+(-1)^n \sqrt{\frac{2\mu}{B(c)}}
\frac{\cos (\wb t )}{\cosh(n\sqrt{\mu  /{V}^{\prime\prime}(c) })}
\label{eq:SABc} ,  \ee
where $\mu =\omega_b^2-4\, {V}^{\prime\prime}(c)$,
$0<\mu\ll 1$.
These solutions are time-periodic (with frequency $\omega_b$)
and their oscillating part is spatially localized.
As $n\rightarrow \pm\infty$ they converge
towards a uniformly stressed static state
(stretched or compressed depending on $c$).
Note that their lower frequency lies inside the
phonon band if ${V}^{\prime\prime}(c) <1$, and
above for ${V}^{\prime\prime}(c) >1$.
The exact solutions approximated by (\ref{eq:SABc}) are
site-centred, i.e. they satisfy
$u_n(t)=u_{-n}(t)$. There exists also exact bond-centred solutions
satisfying $u_n(t)=u_{-n-1} (t+\pi/\omega_{b})$.
In the sequel
we shall denote these solutions as breathers
with an uniform stress.

As we previously mentioned, the numerical study of these
solutions is already contained in section II-B, which applies
to the renormalized displacements $\tilde{x}_n$ satisfying
(\ref{eq:fpuc}).
For the polynomial potential (\ref{eq:V}),
breathers
with an uniform stress have the same
stability properties as the SAB
corresponding to the modified
potential (\ref{modpot}).
In particular,
oscillatory instabilities
are observed for $K_3= 0$
($V$ is even), since the modified
potential $\tilde{V}$ is non-even in this case.
On the contrary, if $K_3 \neq 0$, $K_4 > K_3^2/6$
then the modified potential $\tilde{V}$ is even for
$c=-\frac{K_3}{3K_4}$ (and the conditions
${V}^{\prime\prime}(c) >0$, $B(c)>0$ are realized).
In this case oscillatory instabilities
are not observed for the SAB with
specific static strain $c$.

If $B(c)<0$ and $K_4 >0$,
we numerically obtain large amplitude solutions
of the same type,
whose oscillating parts do not vanish
as $w_b \rightarrow 2\sqrt{{V}^{\prime\prime}(c)}$
(with $w_b > 2\sqrt{{V}^{\prime\prime}(c)}$ ).
This result follows directly from the numerical study of
section III, as well as
the stability properties of these solutions.
These LAB with an uniform stress have been previously
observed by Sandusky and Page
(see \cite{SanduskyP} p.874-876), for the
interaction potential (\ref{eq:V}) with $B<0$.
In their examples $V^{\prime\prime}(c)>1$
(since $c<0$, $K_3 <0$, $K_4 >0$)
and thus the lower frequencies of these
LAB families lie strictly above the phonon band.
This phenomenon can be observed in
fig. 8 of reference \cite{SanduskyP},
which provides frequency vs amplitude plots of such solutions
(note that $c$ may not be constant on each curve).

If $B(c)<0$ and
$\omega_b - 2\sqrt{{V}^{\prime\prime}(c)}<0$
is sufficiently small,
the FPU system has a family of exact
solutions
which can be approximated by
\be u_n (t) \simeq
c+(-1)^n \sqrt{\frac{\mu }{B(c)}}
\tanh\left(\frac{n\sqrt{-\mu }}{\sqrt{2{V}^{\prime\prime}(c)}}\right)
\cos(\wb t)   ,       \label{eq:darkc}\ee
where $\mu =\omega_b^2-4\, {V}^{\prime\prime}(c)<0$,
$|\mu |\ll 1$.
These solutions are time-periodic (with frequency $\omega_b$)
and consists in a dark breather superposed with
a uniformly stressed static state.
Note that their higher frequency lies above the
phonon band if ${V}^{\prime\prime}(c) >1$.
These solutions are site-centred
($u_{-n}(t)=u_{n}(t+\pi/\omega_{b})$), but
their bond-centred analogues also exist.
They have the same
stability properties as dark breathers
in section IV, for the modified
potential (\ref{modpot}).

\section{Conclusions}
In this work we have analyzed both numerically and analytically a rich
variety of nonlinear solutions in FPU lattices. There are four main
results.

On the one hand,
we have numerically explored the range of validity of a recent breathers
(bright or dark) existence proof~\cite{J01} for one-dimensional FPU lattices.
Our computations have been carried out with polynomial interaction potentials
of degree $4$. For $B>0$, we have found that the maximum amplitude of bright
breathers in the relative displacement variables is an approximately linear
function of $\mu^{1/2}$ even far from the phonon band and, in fact, the
approximate expression (\ref{eq:SAB}) describes very well the profile of the
breathers when their amplitudes are large. For $B<0$ and $K_4<0$, we find in
the same way that dark breathers can be very well described by approximate
analytical expressions.

On the other hand, we have found numerically for $B<0$
large amplitude bright
breathers having frequencies arbitrary close to the top of the phonon band,
which are out of range of the local analysis~\cite{J01}. This breather family
exhibits an energy threshold, which is a rarely observed phenomenon in
one-dimensional lattices.

Moreover, with respect to stability properties we have checked that non-even
potentials induce oscillatory instabilities in both bright and dark cases. In
the case of even potentials, the Page mode is linearly stable while the
Sievers-Takeno mode is unstable. We have also found that  bond-centred dark
breathers are unstable. The site-center dark breather is unstable for low
enough frequencies and linearly  stable with an even potential near the edge of
the phonon band.

At last, we have analyzed small amplitude bright and dark breathers
superposed to uniformly stressed static states $x_n=c\, n$.
In particular, the
lowest frequency of bright breathers is inside the phonon band if
$0<{V}^{\prime\prime}(c) <1$, and the highest frequency of dark breathers is
above the phonon band if ${V}^{\prime\prime}(c) >1$. We have obtained local
conditions on $V$ for their existence as well as
approximate analytical expressions.

\begin{acknowledgments}
This work has been supported by the European Union under the RTN project
LOCNET, HPRN--CT--1999--00163. J. Cuevas acknowledges a FPDI grant from la
Junta de Andaluc\'{\i}a. G. James acknowledges stimulating discussions with R.
MacKay.
\end{acknowledgments}

\end{document}